\documentclass[conference]{IEEEtran}
\usepackage[usenames]{color}
\usepackage{amsfonts}
\usepackage{cite}
\usepackage{amssymb,amsthm}
\usepackage{bbm, amsmath}
\usepackage[caption=false]{subfig}
\usepackage{graphicx}
\usepackage{footnote}
\usepackage{algpseudocode}
\usepackage{algorithm}
\usepackage{multirow}
\usepackage{amsmath}
\usepackage{xcolor}
\usepackage{verbatim}
\usepackage{times}
\usepackage{epstopdf}
\usepackage{bm}
\usepackage{color,soul}
\usepackage{tikz}

\usetikzlibrary{patterns}
\usepackage{balance}
\usepackage{pifont}

\title{Evolutionary Game for Consensus Provision in Permissionless Blockchain Networks with Shards}

\author{
\IEEEauthorblockN{Zhengwei Ni\IEEEauthorrefmark{1},
Wenbo Wang\IEEEauthorrefmark{1},
Dong In Kim\IEEEauthorrefmark{2},
Ping Wang\IEEEauthorrefmark{3} and
Dusit Niyato\IEEEauthorrefmark{1}
}
\IEEEauthorblockA{\IEEEauthorrefmark{1}School of Computer Engineering, Nanyang Technological University, Singapore 639798}
\IEEEauthorblockA{\IEEEauthorrefmark{2}School of Information and Communication Engineering, Sungkyunkwan University (SKKU), Suwon, Korea 16419}
\IEEEauthorblockA{\IEEEauthorrefmark{3}Department of Electrical Engineering \& Computer Science, York University, Toronto, Canada ON M3J 1P3}\vspace*{-8mm}}

\IEEEoverridecommandlockouts

\begin{document}
\maketitle
\newtheorem{mydef1}{Lemma}
\newtheorem{mydef2}{Theorem}
\newtheorem{mydef3}{Remark}
\newtheorem{mydef4}{Definition}
\newtheorem{mydef5}{Corollary}
\newtheorem{mydef6}{Proposition}
\begin{abstract}
	With the development of decentralized consensus protocols, permissionless blockchains have been envisioned as a promising enabler for the general-purpose transaction-driven, autonomous systems. However, most of the prevalent blockchain networks are built upon the consensus protocols under the crypto-puzzle framework known as proof-of-work. Such protocols face the inherent problem of transaction-processing bottleneck, as the networks achieve the decentralized consensus for transaction confirmation at the cost of very high latency. In this paper, we study the problem of consensus formation in a system of multiple throughput-scalable blockchains with sharded consensus. Specifically, the protocol design of sharded consensus not only enables parallelizing the process of transaction validation with sub-groups of processors, but also introduces the Byzantine consensus protocols for accelerating the consensus processes. By allowing different blockchains to impose different levels of processing fees and to have different transaction-generating rate, we aim to simulate the multi-service provision eco-systems based on blockchains in real world. We focus on the dynamics of blockchain-selection in the condition of a large population of consensus processors. Hence, we model the evolution of blockchain selection by the individual processors as an evolutionary game. Both the theoretical and the numerical analysis are provided regarding the evolutionary equilibria and the stability of the processors' strategies in a general case.
\end{abstract}
\begin{IEEEkeywords}
Permissionless blockchains, sharding, ELASTICO, evolutionary game
\end{IEEEkeywords}

\section{Introduction}\label{Introduction}
The past decade has witnessed the fast development of the blockchain technologies, especially as the decentralized immutable ledger database (i.e., cryptocurrencies) in the FinTech sector. Most of the studies on blockchains have been focused on the development in cryptocurrencies and the related domains~\cite{7423672}. However, in recent years, more focus about blockchain applications is also put upon the domain of self-organization in general-purpose decentralized systems~\cite{wang2018survey}. Given the decentralized consensus achieved by the blockchain network,  smart contracts~\cite{christidis2016blockchains} are deployed in the form of general-purpose scripts/functions and stored on each consensus node (i.e., processor)\footnote{We use the two terms, i.e., node and processor, interchangeably.} in the network. Following the order prescribed by the consensus about the blockchain states, each smart contract is executed in a replicated manner and thus guarantees to produce a uniform output across the network. Thanks to the technical maturation of smart contracts, blockchains are now envisioned as an enabler for self-organization in wireless networks and decentralized cyber-physical systems. More specifically, existing studies, e.g., autonomous access control~\cite{8269834} and service provision~\cite{8030489}, employ blockchains as an integrator to channel the services upon demands as well as audit the operations of different parties in the system.

\begin{figure}[t]
\centering     %%% not \center
\includegraphics[width=.3\textwidth]{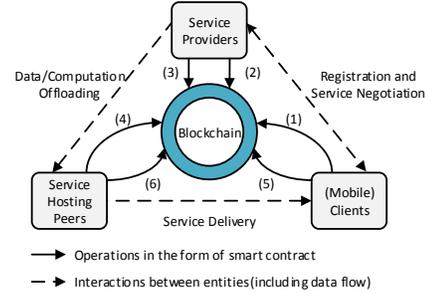}
\caption{A generic framework of blockchain-based self-organization in a service system of three parties. All the deals are settled in a sequence of smart contracts: (1) service requesting by the clients, (2) access granting by the providers, (3) requesting service hosting (e.g., auction for computation/storage/utility offloading) by the providers, (4) settlement of the hosting requests, (5) delivery negotiation between hosting peers and clients and (6) service completion and payment settlement upon proofs of delivery.}
\label{fig_blockchain_intermedeiate}\vspace{-8mm}
\end{figure}

A generic paradigm for blockchain-based self-organization in networking applications is described by Figure~\ref{fig_blockchain_intermedeiate} from the perspective of blockchain users. With the embedded cryptographic functionalities (e.g., asymmetric keys~\cite{wang2018survey}) and the automated transactions based on smart contracts, blockchains are ready to provide the overlaid/virtual channels of secured data/service/payment delivery among trustless parties in the system~\cite{wang2018survey}. As illustrated in Figure~\ref{fig_blockchain_intermedeiate}, this is achieved by encapsulating the controlling rules into smart contracts and the data (e.g., control signals) into blockchain transactions. In particular, when blockchain networks are implemented with permissionless consensus protocols, it is possible to realize the scheme of network management in a purely decentralized manner. Furthermore, when the blockchain maintenance is delegated to groups of nodes with dedicated storage and computing power, we have the blockchain as a Platform as a Service (PaaS) in a similar way to that in the context of cloud computing.

Nevertheless, although permissionless blockchains provide a promising approach to transaction-driven automation for network control problems, most of the existing permissionless consensus protocols are based on Proof-of-Work (PoW)~\cite{7423672} and sacrifice the efficiency, i.e., transaction-processing throughput, for a higher level of consensus security~\cite{wang2018survey}. For example, the popular Ethereum network~\cite{buterin2014ethereum} with a framework of Tuning-complete smart contracts can only support less than 20 Transactions Per Second (TPS). As a result, these blockchains cannot satisfy the low-latency requirements in most of the networking applications and services such as access handing-off between groups of road side units in vehicular-to-infrastructure communication. To guarantee controller response in milliseconds, throughput-scalable protocols such as \textit{ELASTICO}~\cite{Luu:2016:SSP} are proposed to support both open access as in permissionless networks and low latency as in consortium distributed systems~\cite{Vukolic2016}. In brief, the throughput-scalable blockchain networks adopt the PoW-based crypto-puzzle design for node-identity verification and the classical Byzantine Fault-Tolerant (BFT) protocols (e.g., practical BFT~\cite{Castro:2002:PBF}) for distributed transaction ordering. Furthermore, the concept of sharding is adopted from the distributed database~\cite{Croman2016} to enable the parallelization of transaction processing. Thus, the blockchain network is able to increase the TPS as the number of consensus processors increases.

In this paper, we investigate the scenario of general-purpose PaaS based on permissionless blockchains using sharding-based consensus protocols. In particular, we study the problem of consensus provision at the node level for multiple blockchains. Noting that the decentralized processors in a blockchain network are trustless, we assume that the processors are rational (i.e., profit-driven) and non-malicious. Namely, the independent and homogeneous processors participate in the consensus processes of parallel blockchains and dedicate their resources in exchange for the optimal consensus rewards, i.e., the transaction fees collected from the clients. For ease of exposition, we use the ELASTICO protocol~\cite{Luu:2016:SSP} to exemplify the approach of system analysis for consensus participation. Without limiting the blockchains to any specific service provision system, we essentially study a general case of eco-system formation for self-organization with blockchains. Then, by formulating the behaviors of consensus nodes as an evolutionary game, we provide a series of analytical results regarding the equilibrium states and their stability in the evolution of the eco-system.

\section{Preliminaries of Sharded Blockchains}\label{Blockchain_Networks_with_shards}
\subsection{Protocol Fundamentals of ELASTICO}
Blockchain networks with sharding protocol partition the processing processors inside into smaller BFT committees. Each committee processes a disjoint set of transactions, which is called \textit{shard} here. Thus, the tasks of transaction processing are divided into multiple groups and done in parallel. Blockchains with shards overcome the fundamental scalability limits of many popular blockchain systems, e.g., Bitcoin~\cite{Croman2016}. The transaction processing rate is able to scale almost linearly with the number of processors in the network, which ensures that the requirements of real-time resource-access management systems can be met.

Now, we briefly introduce how ELASTICO~\cite{Luu:2016:SSP} works. ELASTICO proceeds in loosely-synchronized epochs, each of which processes a set of transactions. According to \cite{Luu:2016:SSP}, in each epoch, one processor mainly executes 5 procedures:
\begin{enumerate}
	\item The processor is first required to solve PoW puzzles based on the concatenation of a public random seed, its own public key and Internet protocol (IP) address. This procedure allows other processors to verify the identity of the processor. In addition, the processor is randomly assigned to a committee based on the last few bits of its PoW solution. For example, assuming a total number of $2^2=4$ committees, if the last 2 bits of the PoW solution is ``$01$'', the processor will be assigned to committee $2$ if this committee is not full. However, since the target committee to assign to may be full, the processor may need to solve more than one PoW puzzle.
	\item Once a processor is accepted by the network and assigned to a committee, it will discover and establish point-to-point connections with other committee peers following an algorithm of decentralized randomness generation described in~\cite[Section 3.3]{Luu:2016:SSP}.
	\item Then, an authenticated Byzantine agreement protocol, e.g., practical BFT (PBFT)~\cite{Castro:2002:PBF}, is run within a committee to agree on the set of transactions (i.e. shard) allocated to it. Since different committees work in parallel, the network latency only depends on the number of processors in one committee rather than the entire network.
	\item Once an agreement is reached in each committee, all the results will be merged. Then, the final result is broadcast to the network.
	\item Finally, a scheme described in \cite[Section 3.6]{Luu:2016:SSP} is executed by a global committee to randomly generate a new random seed for the next epoch.
\end{enumerate}

\subsection{Average Epoch Time for ELASTICO}
As per the experimental results given by~\cite{Luu:2016:SSP}, the epoch time, i.e., the duration of one epoch, is mainly dominated by two parts, committee formation time and consensus time.

\subsubsection{Average committee formation time}
Committee formation time is the time used for randomly dividing processors into different committees. This time is mainly due to the cost of solving PoW puzzles. We assume that there are totally $n$ processors in the blockchain network. They are divided into $2^s$ committees, with a fixed number of $c$ processors in each committee. Thus, $n=2^sc$. As we mentioned previously, one processor needs to solve more than one PoW puzzle if the originally assigned committee is full. The problem of calculating the total number of the required PoW solutions is equivalent to the extended coupon collector problem~\cite{Double_Dixie_Cup_Problem}. The expected number of PoW solutions is given in \cite[Section 10.1]{Luu:2016:SSP}. When $c$ is fixed, it has a superlinear relationship with $n$, which means that the expected number of PoW puzzles solved by one processor is increasing with $n$. In other words, when the number of processors per committee is fixed, if there are more processors in the blockchain network, one processor is expected to solve more PoW puzzles. In this paper, the expected number of PoW puzzles solved per processor is defined as a continuously differentiable and monotonically increasing function of $n$, $f(n)$, and we have $f(0)=0$. Assume that given a fixed puzzle difficulty, the average time for solving one PoW puzzle is $T$. Then, the average committee formation time can be expressed as $Tf(n)$.

\subsubsection{Average consensus time} The consensus time is determined by the intra-committee agreement for the given shard and the inter-committee agreement for the final result. It is mainly due to the network latency, which is usually caused by the propagation delay of physical links, the forwarding latency of gateways, and the queueing and processing delays of intermediate nodes. In ELASTICO, we can observe that most of communications among processors are limited within the individual committees, so for a given committee size, the time to reach consensus remains almost constant for different network sizes. As \cite[Figure 1]{Luu:2016:SSP} shows, the consensus requires 103 seconds for 400 processors and 110 seconds for 800 processors. Hence, we can think the consensus time is only dependent on the committee size $c$, which is denoted by $g(c)$.

\subsection{Average Reward and Cost in ELASTICO}
In this subsection, we quantitatively model the average reward and cost of one processor per epoch in ELASTICO.
\subsubsection{Average reward}
The processor receives a payment by adding new transaction records into the blockchain. We assume that the transaction records are generated by the users with a rate $\mu$, and the price per transaction is set as $r$. Thus, the average reward of one processor per epoch is $\mu r (Tf(n)+g(c))/n$.
\subsubsection{Average cost}
The cost of one processor is dominated by the energy used for solving PoW puzzles. We assume that the cost of getting one PoW solution is $\varsigma$ on average. Thus, the average cost of one processor per epoch is $\varsigma f(n)$.

\section{System Model and Problem Formulation}
\subsection{Payoff Functions}
We consider $N$ individual processors organizing themselves into $M$ blockchain networks built upon ELASTICO. That is, the processors choose to participate in one of the blockchain networks to receive their revenue. We assume that the processors have identical computing power and the average time for solving one PoW puzzle of fixed difficulty is $T$. We also assume that all blockchain networks adopt the same parameter of committee size $c$. We use the subscript $i$ to denote other parameters for the $i$th blockchain network. Without loss of generality, the index of one blockchain network is determined by $\mu_i r_i$ in a descending order. That is, $\mu_1 r_1\geq \dots\geq\mu_M r_M>0$. The vector of population fractions of the blockchain networks is denoted by $\bm{x}=[x_1,\dots,x_M]^{\top}$, where $[\cdot]^{\top}$ is the notation of transpose. Thus, $\bm{x}$ is in an $(M-1)$-simplex, i.e., $\mathcal{X}=\{\bm{x}\in \mathbb{R}_{+}^{M}: \sum_{i=i}^{M}x_i=1\}$. We call  $\bm{x}$ \textit{state vector} (or \textit{state}) and $\mathcal{X}$ \textit{state space}.

According to Section \ref{Blockchain_Networks_with_shards}, the expected payoff per unit time (i.e., second) of a processor in the $i$th blockchain network can be expressed as
\begin{eqnarray}
u_i(\bm{x})=\frac{\frac{\mu_i r_i (Tf(Nx_i)+g(c))}{Nx_i+\tilde{\tau}}-\varsigma f(Nx_i)}{Tf(Nx_i)+g(c)},\,\, 0\leq x\leq 1,
\end{eqnarray}
where $\tilde{\tau}>0$ can be regarded as the share taken by network operators (e.g., the boosting nodes). By defining $\alpha_i = \mu_i r_i/N$, $\tau = \tilde{\tau}/N$, and $h(x_i) = \varsigma f(Nx_i)/(Tf(Nx_i)+g(c))$, $u_i(\bm{x})$ can be simplified as
\begin{eqnarray}\label{payoff}
u_i(\bm{x})=\frac{\alpha_i}{x_i+\tau}-h(x_i),\,\, 0\leq x\leq 1.
\end{eqnarray}
We can easily obtain that $h(x_i)$ is monotonically increasing with $x_i$ and $h(0)=0$.

\subsection{Dynamical System Formulation}
{In this process,} some processors may switch from one blockchain network to another, causing a change of $\bm{x}$. Since the payoff of the processor is dependent on $\bm{x}$, other processors may also adjust their choice of consensus participation accordingly to choose new blockchain networks. Hence, in this paper, we study population fractions of the blockchain networks as a dynamical system. The state vector at time $t$ is denoted by $\bm{x}(t)=[x_1(t),\dots,x_M(t)]^{\top}$, and we define $\bm{x}_0=\bm{x}(0)$ as the initial state. At time $t$, the rate at which the population fraction of $i$th blockchain network grows is $\frac{dx_i(t)}{dt}$, and we define $\dot{\bm{x}}(t)=\left[\frac{dx_1(t)}{dt},\dots,\frac{dx_M(t)}{dt}\right]^{\top}$. We assume that all the processors are bounded rational and self-interested, so the forces regulating the state vector are from the difference of payoffs. That is, the processors always switch from a blockchain network with low payoff to one with high payoff. Since the payoffs at time $t$ are determined by $\bm{x}(t)$, $\dot{\bm{x}}(t)$ can be described by a function of $\bm{x}(t)$, here defined as $\varphi(\cdot):\mathcal{X}\rightarrow \mathbb{R}^{M}$. Thus, this dynamical system is described by the following ordinary differential equations (ODEs):
\begin{eqnarray}\label{ode}
\dot{\bm{x}}(t)=\varphi(\bm{x}(t)),\,\, \forall t\in \mathbb{R},\,\, i = 1,\dots, M.
\end{eqnarray}
Specially, we adopt the following replicator equations~\cite{weibull1997evolutionary}:
\begin{eqnarray}\label{replicator_equations}
\varphi_i(\bm{x})=x_i\left(u_i(\bm{x})-\bar{u}(\bm{x})\right),
\end{eqnarray}
where $\bar{u}(\bm{x})=\sum_{i=1}^{M}x_iu_i(\bm{x})$, which can be regarded as the average payoff. Notice that here we ignore time $t$ since $\varphi(\cdot)$ is autonomous, that is, does not depend explicitly on time. We can easily find $\sum_{i=1}^{M}\varphi_i(\bm{x})=0$, so that if $\bm{x}_0\in \mathcal{X}$, we always have $\bm{x}(t)\in \mathcal{X}$ for any $t\in \mathbb{R}$.

We are interested in how the vector of population fractions, i.e., state vector, changes with time for different initial states. Usually, it is described by a function $\xi(\cdot,\bm{x}_0): \mathbb{T}\rightarrow \mathcal{X}$, where $\mathbb{T}$ is an open interval containing $t=0$, such that $\xi(0,\bm{x}_0)=\bm{x}_0$, and $\forall t\in \mathbb{T}$,
\begin{eqnarray}
\frac{d}{dt}\xi(t,\bm{x}_0)=\varphi\left(\xi(t,\bm{x}_0)\right).
\end{eqnarray}
The function $\xi(\cdot,\bm{x}_0)$ is called a \textit{solution} of \eqref{ode}.

\section{Analysis of the Game}
\subsection{Uniqueness of Solutions for Different Initial Points}
In this subsection, we show that $\forall \bm{x}_0\in \mathcal{X}$, we have a unique solution $\xi(\cdot,\bm{x}_0)$. It means that once the initial state is determined, how the population fractions evolve over time is totally determined. It is stated in the following theorem.
\begin{mydef2}\label{Uniqueness}
For $\varphi(\cdot):\mathcal{X}\rightarrow \mathbb{R}^{M}$ described in \eqref{replicator_equations} and $\forall \bm{x}_0\in \mathcal{X}$, the system \eqref{ode} has a unique solution.
\end{mydef2}
\begin{proof}
	We can obtain that
	\begin{eqnarray}
	&{}&\frac{\partial \varphi_i(\bm{x})}{\partial x_i}=(1-2x_i)\left(\frac{\alpha_i}{x_i+\tau}-h(x_i)\right)\nonumber\\&-&(x_i-x_i^2)\left(\frac{\alpha_i}{(x_i+\tau)^2}+\frac{dh(x)}{dx}\bigg|_{x=x_i}\right)\nonumber\\&-&\sum_{j=1,j\neq i}^{M}x_j\left(\frac{\alpha_j}{x_j+\tau}-h(x_j)\right),\label{i_i}\\
	&{}&\frac{\partial \varphi_i(\bm{x})}{\partial x_j}=-x_i\left(\frac{\alpha_j}{x_j+\tau}-h(x_j)\right)\nonumber\\&+&x_ix_j\left(\frac{\alpha_j}{(x_j+\tau)^2}+\frac{dh(x)}{dx}\bigg|_{x=x_j}\right).\label{i_j}
	\end{eqnarray}
	(\ref{i_i}) and (\ref{i_j}) indicate that $\frac{\partial \varphi_i(\bm{x})}{\partial x_i}$ and $\frac{\partial \varphi_i(\bm{x})}{\partial x_j}$ exist and are continuous in $\mathcal{X}$. Hence, $\varphi(\bm{x})$ is Lipschitz continuous in $\mathcal{X}$. By the Picard-Lindel{\"o}f theorem \cite[Theorem 6.1]{weibull1997evolutionary}, we obtain Theorem~\ref{Uniqueness}.
\end{proof}

\subsection{Existence of Equilibria}\label{existence_of_equilibria}
Mathematically, an \textit{equilibrium} (a.k.a., \textit{rest point} or \textit{critical point}) under a solution mapping $\xi$ is a state vector $\bm{x}\in\mathcal{X}$ such that $\xi(t,\bm{x})= \bm{x}$ for all $t\in \mathbb{R}$ \cite[Definition 6.4]{weibull1997evolutionary}. In our model, it means that if the vector of population fractions is at an equilibrium, this population distribution will remain the same, which implies that there are no ``job-hoppings'' in the blockchain networks. In addition, according to \cite[Proposition 6.3]{weibull1997evolutionary}, if the vector of population fractions finally converges over time, it will converge to an equilibrium.

Now, we are ready to give all the possible equilibria. Based on whether there are any processor, we can divide the considered blockchain networks into two specific sets, i.e., the \textit{working blockchain set} $\mathcal{W}=\{i:x_i>0\}$ and the \textit{resting blockchain set} $\bar{\mathcal{W}}=\{i:x_i=0\}$. A state is an equilibrium if and only if $\varphi(\bm{x})$ vanishes at this state. For a given $\mathcal{W}=\{{i_1},\dots,{i_{|\mathcal{W}|}}\}$, it means that
\begin{eqnarray}
u_{i_1}(\bm{x})=\dots=u_{i_{|\mathcal{W}|}}(\bm{x}).
\end{eqnarray}
Theoretically, there are totally $2^M-1$ possible $\mathcal{W}$. However, for some values of $\mathcal{W}$, there may not be any equilibria. Now we give the conditions that a given $\mathcal{W}$ has at least one equilibrium.

Consider a field $\mathcal{K}=\{(a,b):\,a,b>0,\,\frac{a}{1+\tau}-h(1)\leq b\leq\frac{a}{\tau}\}\subset \mathbb{R}^2$, and a function $K(\cdot):\mathcal{K}\rightarrow [0,1]$ such that $K(\hat{a},\hat{b})$ is the solution for the equation
\begin{eqnarray}\label{K}
\frac{\hat{a}}{x+\tau}-h(x)=\hat{b}.
\end{eqnarray}
Notice that when $K(\hat{a},\hat{b})$ is continuous and monotonically increasing with $\hat{a}$ and decreasing with $\hat{b}$.
Then, we give the following theorem.
\begin{mydef2}\label{equilibrium_existence}
For a given set of working blockchains $\mathcal{W}=\{{i_1},\dots,{i_{|\mathcal{W}|}}\}$, if $\frac{\alpha_{i_1}}{1+\tau}-h(1)<\frac{\alpha_{i_{|\mathcal{W}|}}}{\tau}$ and $\sum_{j=1}^{|\mathcal{W}|-1}K\left(\alpha_{i_j},\frac{\alpha_{i_{|\mathcal{W}|}}}{\tau}\right)<1$, a unique equilibrium exists. Otherwise, there is no equilibrium.
\end{mydef2}
\begin{proof} First, we show if $\frac{\alpha_{i_1}}{1+\tau}-h(1)<\frac{\alpha_{i_{|\mathcal{W}|}}}{\tau}$ and $\sum_{j=1}^{|\mathcal{W}|-1}K\left(\alpha_{i_j},\frac{\alpha_{i_{|\mathcal{W}|}}}{\tau}\right)<1$, a unique equilibrium exists.
	Since $\alpha_{i_1}\geq\dots\geq \alpha_{i_{|\mathcal{W}|}}$, if $\frac{\alpha_{i_1}}{1+\tau}-h(1)<\frac{\alpha_{i_{|\mathcal{W}|}}}{\tau}$, we must have $\left(\alpha_{i_j},\frac{\alpha_{i_{|\mathcal{W}|}}}{\tau}\right)\in \mathcal{K}$, $j=1,\dots, |\mathcal{W}|-1$, and
	\begin{eqnarray}
	0\leq K\left(\alpha_{i_j},\frac{\alpha_{i_{|\mathcal{W}|}}}{\tau}\right)<1,\,\, j=1,\dots,|\mathcal{W}|-1.
	\end{eqnarray}
	Notice that $K\left(\alpha_{i_{|\mathcal{W}|}},\frac{\alpha_{i_{|\mathcal{W}|}}}{\tau}\right)=0$. In addition, since for $j=2,\dots,|\mathcal{W}|$,
	\begin{eqnarray}
	\frac{\alpha_{i_j}}{1+\tau}-h(1)\leq\frac{\alpha_{i_1}}{1+\tau}-h(1)<\frac{\alpha_{i_{|\mathcal{W}|}}}{\tau}\leq\frac{\alpha_{i_j}}{\tau},
	\end{eqnarray}
	we have
	\begin{eqnarray}
	0< K\left(\alpha_{i_j},\frac{\alpha_{i_1}}{1+\tau}-h(1)\right) \leq 1,\,\, j=2,\dots,|\mathcal{W}|.
	\end{eqnarray}
	Obviously the following holds
	\begin{eqnarray}
	K\left(\alpha_{i_1},\frac{\alpha_{i_1}}{1+\tau}-h(1)\right) = 1.
	\end{eqnarray}
	Thus,
	\begin{eqnarray}
	\sum_{j=1}^{|\mathcal{W}|}K\left(\alpha_{i_j},\frac{\alpha_{i_1}}{1+\tau}-h(1)\right)>1.
	\end{eqnarray}
Since  $\sum_{j=1}^{|\mathcal{W}|}K\left(\alpha_{i_j},\frac{\alpha_{i_{|\mathcal{W}|}}}{\tau}\right)<1$ and $K(\hat{a},\hat{b})$ is continuous and monotonically decreasing with $\hat{b}$, there must exist a unique $\bar{b}\in\left(\frac{\alpha_{i_1}}{1+\tau}-h(1),\frac{\alpha_{i_{|\mathcal{W}|}}}{\tau}\right)$ such that
\begin{eqnarray}
\sum_{j=1}^{|\mathcal{W}|}K\left(\alpha_{i_j},\bar{b}\right)=1,
\end{eqnarray}
and the population fraction of the $i_j$th blockchain network is indeed $K\left(\alpha_{i_j},\bar{b}\right)$.

When $\frac{\alpha_{i_1}}{1+\tau}-h(1)\geq\frac{\alpha_{i_{|\mathcal{W}|}}}{\tau}$. $\forall x_{i_1},x_{i_{|\mathcal{W}|}}\in(0,1]$, we can easily obtain that
\begin{eqnarray}
&{}&\frac{\alpha_{i_{|\mathcal{W}|}}}{x_{i_{|\mathcal{W}|}}+\tau}-h\left(x_{i_{|\mathcal{W}|}}\right)<\frac{\alpha_{i_{|\mathcal{W}|}}}{\tau}\nonumber\\&\leq&\frac{\alpha_{i_1}}{1+\tau}-h(1)\leq \frac{\alpha_{i_1}}{x_{i_1}+\tau}-h(x_{i_1}).
\end{eqnarray}
Hence, there are no equilibria in this case.

Finally, it is obvious that when $\frac{\alpha_{i_1}}{1+\tau}-h(1)<\frac{\alpha_{i_{|\mathcal{W}|}}}{\tau}$ and $\sum_{j=1}^{|\mathcal{W}|-1}K\left(\alpha_{i_j},\frac{\alpha_{i_{|\mathcal{W}|}}}{\tau}\right)\geq1$, $u_{i_1}(\bm{x})=\dots=u_{i_{|\mathcal{W}|}}(\bm{x})>0$ will lead to $\sum_{j=1}^{|\mathcal{W}|}x_{i_j}>1$. Hence, there are no equilibria in this case.
\end{proof}

\subsection{Asymptotic Stability of the Equilibria}
Among all the equilibria, we are especially interested in those which are ``robust''. That means, all sufficiently small perturbations of the equilibrium induce a backward movement. The mathematical definition of asymptotic stability in a dynamical system can be found in \cite[Definitionn 6.5]{weibull1997evolutionary}.

We assume that $\bm{x}^*$ is an equilibrium and let $J_{\bm{x}^*}^{\varphi}$ be the $M\times M$ Jacobian matrix of $\varphi(\cdot)$ at the state $\bm{x}^*$. Then, $J_{\bm{x}^*}^{\varphi}$ can be expressed as
\begin{eqnarray}
J_{\bm{x}^*}^{\varphi}=\left.\left[\begin{array}{llll}
\frac{\partial \varphi_1(\bm{x})}{\partial x_1} & \frac{\partial \varphi_1(\bm{x})}{\partial x_2} & \dots &\frac{\partial \varphi_1(\bm{x})}{\partial x_M}\\\frac{\partial \varphi_2(\bm{x})}{\partial x_1} & \frac{\partial \varphi_2(\bm{x})}{\partial x_2} & \dots &\frac{\partial \varphi_2(\bm{x})}{\partial x_M}\\\vdots & \vdots & \ddots & \vdots\\\frac{\partial \varphi_M(\bm{x})}{\partial x_1} & \frac{\partial \varphi_M(\bm{x})}{\partial x_2} & \dots &\frac{\partial \varphi_M(\bm{x})}{\partial x_M}
\end{array}\right]\right|_{\bm{x}=\bm{x^*}},
\end{eqnarray}
where $\frac{\partial \varphi_i(\bm{x})}{\partial x_i}$ and $\frac{\partial \varphi_i(\bm{x})}{\partial x_j}$, $j\neq i$ are given in \eqref{i_i} and \eqref{i_j}, respectively. According to \cite[Theorem 8.4.3]{lebovitz1999ordinary}, $\bm{x^*}$ is asymptotically stable if the real part of every eigenvalue of $J_{\bm{x}^*}^{\varphi}$ is negative, and it is unstable if any eigenvalue of $J_{\bm{x}^*}^{\varphi}$ has a positive real part.
Based on our observations of $J_{\bm{x}^*}^{\varphi}$, we can obtain the following lemma.
\begin{mydef1}\label{eigenvalue}
For a given set of working blockchains $\mathcal{W}=\{{i_1},\dots,{i_{|\mathcal{W}|}}\}$ and its corresponding equilibrium $\bm{x}_{\mathcal{W}}^*$, $\forall k\in\bar{\mathcal{W}}$, $\lambda_{k}=\frac{\alpha_k}{\tau}-\frac{\alpha_{i_{|\mathcal{W}|}}}{x_{i_{|\mathcal{W}|}}+\tau}+h\left(x_{i_{|\mathcal{W}|}}\right)$ is one eigenvalue of $J_{\bm{x}^*}^{\varphi}$.
\end{mydef1}
\begin{proof}
	From \eqref{i_i} and \eqref{i_j}, we can obtain that $\forall k\in\bar{\mathcal{W}}$,
	\begin{eqnarray}\label{partial_1}
	&{}&\left.\frac{\partial \varphi_k(\bm{x})}{\partial x_k}\right|_{x_k=0}\nonumber\\&=&\frac{\alpha_k}{\tau}-\sum_{j=1,j\neq k}^{M}x_j\left(\frac{\alpha_j}{x_j+\tau}-h(x_j)\right)\nonumber\\&=&\frac{\alpha_k}{\tau}-\left(\frac{\alpha_{i_{|\mathcal{W}|}}}{x_{i_{|\mathcal{W}|}}+\tau}-h\left(x_{i_{|\mathcal{W}|}}\right)\right)=\lambda_k.
	\end{eqnarray}
In addition, when $x_k=0$, $\forall \tilde{k}\neq k$, we have
	\begin{eqnarray}\label{partial_2}
	\frac{\partial \varphi_k(\bm{x})}{\partial x_{\tilde{k}}}=0.
	\end{eqnarray}
	We can observe that the matrix $J_{\bm{x}_{\mathcal{W}}^*}^{\varphi}-\lambda_k I_{M\times M}$ has its elements of $k$th row all equal to 0, where $I_{M\times M}$ is the $M\times M$ identity matrix. Thus, ${\rm det}\left(J_{\bm{x}_{\mathcal{W}}^*}^{\varphi}-\lambda_k I_{M\times M}\right)=0$, where ${\rm det}(\cdot)$ is the determinant. Hence, $\lambda_k$ is an eigenvalue of $J_{\bm{x}_{\mathcal{W}}^*}^{\varphi}$.
\end{proof}

Based on the Lemma \ref{eigenvalue}, we can further obtain Theorems~\ref{stability_1}.
\begin{mydef2}\label{stability_1}
	For a given set of working blockchains $\mathcal{W}=\{{i_1},\dots,{i_{|\mathcal{W}|}}\}$ and its corresponding equilibrium $\bm{x}_{\mathcal{W}}^*$, if $\exists k\in \bar{\mathcal{W}}$, and $\lambda_k\geq0$, this equilibrium is not asymptotically stable.
\end{mydef2}
\begin{proof}
	From Lemma \ref{eigenvalue}, we know that $\lambda_k$ is an eigenvalue of $J_{\bm{x}_{\mathcal{W}}^*}^{\varphi}$. According to \cite[Theorem 8.4.3]{lebovitz1999ordinary}, $J_{\bm{x}_{\mathcal{W}}^*}^{\varphi}$ has at least one non-negative eigenvalue, so $\bm{x}_{\mathcal{W}}^*$ is not asymptotically stable.
\end{proof}

Then, we can obtain the following corollaries.
\begin{mydef5}\label{c1}
	A set of working blockchains $\mathcal{W}=\{{i_1},\dots,{i_{|\mathcal{W}|}}\}$ with $|\mathcal{W}|=w$ has a corresponding equilibrium. If $\mathcal{W}\neq [1,w]\cap \mathbb{N}$, this equilibrium is not asymptotically stable.
\end{mydef5}
\begin{proof}
	Since $\mathcal{W}\neq [1,w]\cap \mathbb{N}$, there must exist a positive integer $l$ such that $l\leq w$ and $l\in \bar{\mathcal{W}}$. Clearly, $\lambda_l\geq 0$. According to Theorem \ref{stability_1}, we have this corollary.
\end{proof}
\begin{mydef5}\label{c2}
	If the set of working blockchains $[1,w]\cap \mathbb{N}$ has a corresponding equilibrium, for any positive integer $r$ such that $r<w$, the set of working blockchains $[1,r]\cap \mathbb{N}$ must have a corresponding equilibrium, which is not asymptotically stable.
\end{mydef5}
\begin{proof}
	It is easy to obtain
	\begin{eqnarray}
	\frac{\alpha_{1}}{1+\tau}-h(1)<\frac{\alpha_{w}}{\tau}\leq\frac{\alpha_{r}}{\tau},
	\end{eqnarray}
	and
	\begin{eqnarray}
	&{}&\sum_{j=1}^{r-1}K\left(\alpha_{j},\frac{\alpha_{r}}{\tau}\right)\leq \sum_{j=1}^{r-1}K\left(\alpha_{j},\frac{\alpha_{w}}{\tau}\right)\nonumber\\&\leq&\sum_{j=1}^{w-1}K\left(\alpha_{j},\frac{\alpha_{w}}{\tau}\right)<1.
	\end{eqnarray}
	Hence, according to Theorem \ref{equilibrium_existence}, the set of working blockchains $[1,r]\cap \mathbb{N}$ must have an equilibrium.

	We denote this equilibrium by $\bm{x}_{[1,r]\cap \mathbb{N}}^*$, with Jacobian matrix $J_{\bm{x}_{[1,r]\cap \mathbb{N}}^*}^{\varphi}$. Let $\sum_{j=1}^{r}K(\alpha_j,\bar{b}_r)=1$ and $\sum_{j=1}^{w}K(\alpha_j,\bar{b}_w)=1$. Clearly, $\bar{b}_r< \bar{b}_w$. Since $w\notin [1,r]\cap \mathbb{N}$, according to Lemma \ref{eigenvalue}, $\frac{\alpha_w}{\tau}-\bar{b}_r$ is an eigenvalue of $J_{\bm{x}_{[1,r]\cap \mathbb{N}}^*}^{\varphi}$. Since $\frac{\alpha_w}{\tau}>\bar{b}_w> \bar{b}_r$, according to Theorem \ref{stability_1}, $\bm{x}_{[1,r]\cap \mathbb{N}}^*$ is not asymptotically stable.
\end{proof}
\begin{mydef2}\label{t4}
	Let $w^* = \max\{w:\frac{\alpha_{1}}{1+\tau}-h(1)<\frac{\alpha_{w}}{\tau}, \sum_{j=1}^{w-1}K\left(\alpha_{j},\frac{\alpha_{w}}{\tau}\right)<1\}$. Then, only the corresponding equilibrium of the set of working blockchains $[1,w^*]\cap \mathbb{N}$ has probability to be asymptotically stable.
\end{mydef2}
\begin{proof}
	According to Corollaries \ref{c1} and \ref{c2}, all the other equilibria are not asymptotically stable. Then, we must have Theorem \ref{t4}.
\end{proof}

\section{Performance Evaluation}
In this section, we provide the numerical analysis of the population dynamics of ELASTICO. We assume four blockchain networks, for which $\alpha_1 = 0.7$, $\alpha_2 = 0.5$, $\alpha_3 = 0.3$, and $\alpha_4 = 0.1$. In addition, we set $\tau = 0.01$ and $h(x) = \ln(1+x)$. In this condition, from our analysis in Section~\ref{existence_of_equilibria}, we can easily obtain $\bm{x}_{e1} = [0.4225,	0.3148,	0.1975,	0.0652]^{\top}$ and $\bm{x}_{e2} = [0.4499,	0.3369,	0.2132,	0]^{\top}$ are two equilibria. By calculating the eigenvalues of Jacobian matrix of $\varphi(\cdot)$ at the state $\bm{x}_{e1}$, we can find that $\bm{x}_{e1}$ is asymptotically stable. Meanwhile, according to Corollary~\ref{c2}, we can see that $\bm{x}_{e2}$ is not asymptotically stable.

\begin{figure}[!t]
	\begin{center}
		\hspace*{-5mm}
		\includegraphics[width=.43\textwidth]{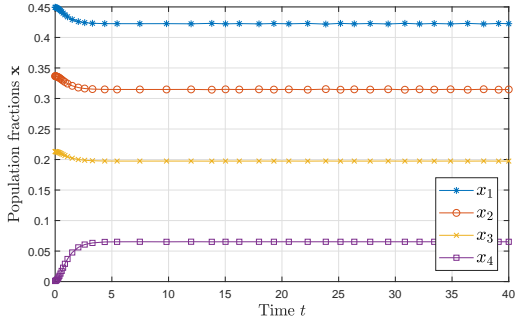}\vspace{-5mm}
	\end{center}
	\caption{The population dynamics for four blockchain networks, from the initial point $\bm{x_0}=[0.4498,	0.3369,	0.2132,	0.001]^{\top}$.}\label{population_dynamics}
\end{figure}

\begin{figure}[!t]
	\begin{center}
		\hspace*{-5mm}
		\includegraphics[width=.43\textwidth]{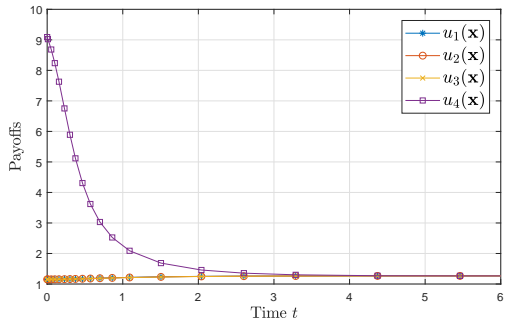}
	\end{center}
	\caption{The payoff dynamics for four blockchain networks, from the initial point $\bm{x_0}=[0.4498,	0.3369,	0.2132,	0.001]^{\top}$.}\label{payoff_dynamics}\vspace{-5mm}
\end{figure}

We show the population dynamics from an initial point $\bm{x_0}=[0.4498,	0.3369,	0.2132,	0.001]^{\top}$, which is a very small deviation from $\bm{x}_{e2}$. From Figure~\ref{population_dynamics}, we can observe that, instead of moving backward to its neighboring equilibrium $\bm{x}_{e2}$, the state moves away from this area and settles down over time toward the new equilibrium $\bm{x}_{e1}$. This observation coincides with our analysis given in the previous paragraph.
Then, in Figure~\ref{payoff_dynamics}, we show the dynamics of payoff for each blockchain network. We can observe that at the beginning, since there are only a few processors in the fourth blockchain, the payoff per second of one processor is much higher than other three blockchains. However, as more and more processors move to the fourth blockchain, the payoff for the fourth blockchain decreases while those for the other three blockchains increase. Finally, they reach an equilibrium and all the blockchains have the same expected payoff.

\begin{figure}[!t]
	\begin{center}
		\hspace*{-5mm}
		\includegraphics[width=.43\textwidth]{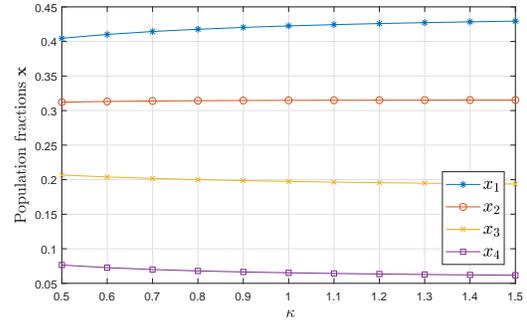}
	\end{center}
	\caption{The asymptotically stable equilibrium for $\alpha_1 = 0.7\kappa$, $\alpha_2 = 0.5\kappa$, $\alpha_3 = 0.3\kappa$, and $\alpha_4 = 0.1\kappa$, where $\kappa$ changes from $0.5$ to $1.5$.}\label{equilibriumvsprices}\vspace{-5mm}
\end{figure}

Finally, we show how the asymptotically stable equilibrium changes with the prices. We set $\alpha_1 = 0.7\kappa$, $\alpha_2 = 0.5\kappa$, $\alpha_3 = 0.3\kappa$, and $\alpha_4 = 0.1\kappa$, where $\kappa$ varies from $0.5$ to $1.5$. Figure \ref{equilibriumvsprices} shows that, as the prices increase, the population fractions of blockchain networks with higher prices increase while those of blockchain networks with lower prices decrease.

\section{Conclusion}
In this paper, we have investigated the process of eco-system formation for a large population of consensus nodes to join one consensus process from multiple permissionless, sharded blockchains. In particular, we have considered the scenario of multiple blockchains adopting the ELASTICO protocol, which combines the proof-of-work puzzle and the Byzantine fault-tolerant agreement protocol to achieve both open access and low transaction-processing latency. We have considered a general scenario where different blockchains may provide different transaction fees and have different transaction-generating rates. We have studied the blockchain-selection behaviors of the independent, bounded-rational consensus nodes with identical computational power. The behaviors of blockchain selection by the consensus nodes have been formulated as an evolutionary game based on replicator dynamics. We have provided a series of analytical and numerical results, which reveal the consensus-formation mechanism in a permissionless network of blockchains for multiple service provision.

\bibliographystyle{IEEEtran}
\bibliography{IEEEabrv,mybibfile}

\begin{thebibliography}{10}
\providecommand{\url}[1]{#1}
\csname url@samestyle\endcsname
\providecommand{\newblock}{\relax}
\providecommand{\bibinfo}[2]{#2}
\providecommand{\BIBentrySTDinterwordspacing}{\spaceskip=0pt\relax}
\providecommand{\BIBentryALTinterwordstretchfactor}{4}
\providecommand{\BIBentryALTinterwordspacing}{\spaceskip=\fontdimen2\font plus
\BIBentryALTinterwordstretchfactor\fontdimen3\font minus
  \fontdimen4\font\relax}
\providecommand{\BIBforeignlanguage}[2]{{%
\expandafter\ifx\csname l@#1\endcsname\relax
\typeout{** WARNING: IEEEtran.bst: No hyphenation pattern has been}%
\typeout{** loaded for the language `#1'. Using the pattern for}%
\typeout{** the default language instead.}%
\else
\language=\csname l@#1\endcsname
\fi
#2}}
\providecommand{\BIBdecl}{\relax}
\BIBdecl

\bibitem{7423672}
F.~Tschorsch and B.~Scheuermann, ``Bitcoin and beyond: A technical survey on
  decentralized digital currencies,'' \emph{IEEE Communications Surveys
  Tutorials}, vol.~18, no.~3, pp. 2084--2123, third quarter 2016.

\bibitem{wang2018survey}
W.~Wang, D.~T. Hoang, Z.~Xiong, D.~Niyato, P.~Wang, P.~Hu, and Y.~Wen, ``A
  survey on consensus mechanisms and mining management in blockchain
  networks,'' \emph{arXiv preprint arXiv:1805.02707}, 2018.

\bibitem{christidis2016blockchains}
K.~Christidis and M.~Devetsikiotis, ``Blockchains and smart contracts for the
  internet of things,'' \emph{IEEE Access}, vol.~4, pp. 2292--2303, May 2016.

\bibitem{8269834}
K.~Kotobi and S.~G. Bilen, ``Secure blockchains for dynamic spectrum access: A
  decentralized database in moving cognitive radio networks enhances security
  and user access,'' \emph{IEEE Vehicular Technology Magazine}, vol.~13, no.~1,
  pp. 32--39, Mar. 2018.

\bibitem{8030489}
N.~Herbaut and N.~Negru, ``A model for collaborative blockchain-based video
  delivery relying on advanced network services chains,'' \emph{IEEE
  Communications Magazine}, vol.~55, no.~9, pp. 70--76, 2017.

\bibitem{buterin2014ethereum}
\BIBentryALTinterwordspacing
V.~Buterin, ``Ethereum: A next-generation smart contract and decentralized
  application platform,'' Ethereum Foundation, Tech. Rep., 2014. [Online].
  Available: \url{https://github.com/ethereum/wiki/wiki/White-Paper}
\BIBentrySTDinterwordspacing

\bibitem{Luu:2016:SSP}
L.~Luu, V.~Narayanan, C.~Zheng, K.~Baweja, S.~Gilbert, and P.~Saxena, ``A
  secure sharding protocol for open blockchains,'' in \emph{Proceedings of the
  2016 ACM SIGSAC Conference on Computer and Communications Security}.\hskip
  1em plus 0.5em minus 0.4em\relax New York, NY, USA: ACM, 2016, pp. 17--30.

\bibitem{Vukolic2016}
M.~Vukoli{\'{c}}, ``The quest for scalable blockchain fabric: Proof-of-work vs.
  bft replication,'' in \emph{Open Problems in Network Security: IFIP WG 11.4
  International Workshop}, Zurich, Switzerland, Oct. 2015, pp. 112--125.

\bibitem{Castro:2002:PBF}
M.~Castro and B.~Liskov, ``Practical byzantine fault tolerance and proactive
  recovery,'' \emph{ACM Trans. Comput. Syst.}, vol.~20, no.~4, pp. 398--461,
  Nov. 2002.

\bibitem{Croman2016}
K.~Croman, C.~Decker, I.~Eyal, A.~E. Gencer, A.~Juels, A.~Kosba, A.~Miller,
  P.~Saxena, E.~Shi, E.~G{\"u}n~Sirer, D.~Song, and R.~Wattenhofer, ``On
  scaling decentralized blockchains,'' in \emph{Financial Cryptography and Data
  Security: International Workshops on BITCOIN, VOTING and WAHC}, Christ
  Church, Barbados, Feb. 2016, pp. 106--125.

\bibitem{Double_Dixie_Cup_Problem}
D.~J. Newman, ``The double dixie cup problem,'' \emph{The American Mathematical
  Monthly}, vol.~67, no.~1, pp. 58--61, 1960.

\bibitem{weibull1997evolutionary}
J.~W. Weibull, \emph{Evolutionary game theory}.\hskip 1em plus 0.5em minus
  0.4em\relax MIT press, 1997.

\bibitem{lebovitz1999ordinary}
N.~Lebovitz, \emph{Ordinary Differential Equations}, ser. Mathematics
  Series.\hskip 1em plus 0.5em minus 0.4em\relax Brooks/Cole, 1999.

\end{thebibliography}

\end{document}